# Efficient Interaction of Heralded X-ray Photons with a Beam Splitter

E. Strizhevsky[1], D. Borodin[1], A. Schori[1,2], S. Francoual[3], R. Röhlsberger[3], and S. Shwartz[1*]

[1]*Physics Department and Institute of Nanotechnology, Bar-Ilan University, Ramat Gan, 5290002 Israel*

[2]*PULSE Institute, SLAC National Accelerator Laboratory, Menlo Park, California 94025, USA*

[3]*Deutsches Elektronen-Synchrotron DESY, Notkestrasse 85, D-22607 Hamburg, Germany*

We report the experimental demonstration of efficient interaction of multi-kilo-electron–Volt heralded x-ray photons with a beam splitter. The measured heralded photon rate at the outputs of the beam splitter is about 0.01 counts/s which is comparable to the rate in the absence of the beam splitter. We use this beam splitter together with photon number and photon energy resolving detectors to show directly that single x-ray photons cannot split. Our experiment demonstrates the major advantage of x-rays for quantum optics – the possibility to observe experimental results with high fidelity and with negligible background.

Beam splitters, which are devices that split electromagnetic radiation, are among the most important optical components for quantum optics [1–4]. They are the essential components in almost any experiment aiming at the study of fundamental quantum optics and serve as the building blocks for almost any optical quantum technology. Indeed, seminal works showing the quantum nature of light using beam splitters include, for example, the Hong-Ou-Mandel effect [5], interaction free measurements [6,7], interaction of single photons with beam splitter [8], and the generation and measurements of entanglement [9] and NOON states [10].

The extension of quantum optics into the x-ray regime would have a tremendous impact [11]. Concepts of quantum optics can lead, for instance, to significant reduction of the dose used for imaging [12–14] and to the enhancement of the sensitivity [15], and the signal-to-noise ratio (SNR) of measurements [16–21]. Furthermore, the availability of commercial detectors that reach nearly 100% efficiency with low dark current and real capabilities of photon number resolving over a very broad spectral range is extremely appealing for tests of basic concepts in quantum optics [11,22].

However, despite the pronounced potential, the utilization of beam splitters for x-ray quantum optics has never been demonstrated. The main challenge is finding beam splitters than can facilitate the broad spectral and angular widths of the generated quantum states of x-ray radiation.

The two potential sources for the generation of nonclassical forms of radiation in the x-ray regime are radioactive sources with a cascade scheme that leads to the emission of two simultaneous photons and spontaneous parametric down-conversion (SPDC) in which pairs of entangled photons are generated [23]. The first has been demonstrated with Mossbauer nuclei [24,25] but although exhibits a very narrow spectral range, the emission is in all directions, thus it is challenging to collect a sufficient portion of the emerging photons. In SPDC the spectral width of the generated photons is in the multi-keV range and the angular width is several degrees [26–28]. However, in most cases, x-ray optics relies on either Bragg scattering or on reflection from surfaces [29]. For Bragg scattering from crystals the typical values for the angular acceptance and spectral width are a few mdeg and eV, respectively. Accordingly, those devices cannot render the interaction with the broad SPDC signal efficient. Reflections from surfaces works well only at grazing incident angles and cannot be used either. The two conceivable candidates are mosaic crystals [29,30] and Nano-scale multilayer periodic structures [31]. Both can be designed to support acceptance angles in the several degrees range and with spectral lineshapes exceeding several hundred electron-Volts. However, the parameters have to be selected carefully to maintain high simultaneous reflectance and transmittance.

In this letter we describe how to utilize broad spectral and angular bandwidth x-ray beam splitters for x-ray quantum optics. We use the broadband heralded photons generated by SPDC as a quantum state and show that their interaction with the beam splitter is efficient by comparing the coincidence rates before and after the beam splitter. Our approach to realize efficient interaction is to use a mosaic crystal as a Bragg beam splitter with a wide rocking curve width and to choose its angular dispersion to match the angular dispersion of the photon pairs. We prefer the mosaic crystal over multilayers to avoid the loss in the substrate of those devices. We employ the beam splitter to demonstrate directly and without background noise that a single x-ray photon is an unsplittable quantum despite the unavoidable loss in the system. This is a direct confirmation of the prediction of Barnett et al. who considered a quantum theory for the interaction with lossy beam splitters [32].

The setup we use in this work and that is based on the standard scheme for generating and detecting heralded photons [26] is depicted in Fig. 1(a). The process of SPDC is used to generate photon pairs in the nonlinear medium. Since the photons are always generated in pairs, once we detect one photon, we know with certainty that the second photon exists. This second photon, is called heralded, and the heralded photons exhibit all the properties of single photons including sub-Poisson statistics, which is a clear distinction from classical radiation. Of interest to the present work is that a true single photon cannot split even when it interacts with a beam splitter in contrast to classical beams. This is manifested in the coincidence measurements between the two output ports of the beam splitter, which are null when using ideal single photon sources and detectors.

In our scheme a pump beam at $\hbar\omega_\text{p} = 21$ keV hits upon a nonlinear crystal, which is a 4 mm × 4 mm × 0.8 mm diamond crystal, to generate photon pairs both at central photon energy of 10.5 keV by SPDC (see Supplemental Material [33] for further details). The reciprocal lattice vector normal to the C(660) atomic planes is used for phase matching, and the detectors are silicon drift detectors (SDDs) with an active area of 25 mm$^2$ [34]. We use a Highly Ordered Pyrolytic Graphite (HOPG) for the beam splitter with the dimensions of 20 mm × 20 mm × 0.7 mm. We use its (002) atomic planes where the Bragg angle is 10.1° for the central photon energy (10.5 keV). For each photon pair, one photon at $\hbar\omega_{Trig}$ is denoted as the trigger photon and is measured directly by the detector $D_{Trig}$. The second photon at $\hbar\omega_{Heral}$ is the heralded photon and it hits upon a beam splitter and collected by either $D_{Ref}$ or $D_{Trans,}$, which are the detectors for the reflected and transmitted beams, respectively.

To find parameters that can support high-efficient beam splitter interaction with single photons we calculate numerically the rate of the heralded photons by using the second order Glauber correlation function [26] where we consider a Gaussian function to model the reflection coefficient of the beam splitter (see Supplemental Material [33]).

We show below that the important parameter is the Bragg angle that for a given input wavelength is determined by the lattice interplanar spacing, thus can serve as a guide for the selection of the material and the crystallographic orientation of the beam splitter. In Fig. 1(b) we show the theoretical dependence of the heralded photon count rate on the Bragg angle of the beam splitter for the experimental parameters described above. From Fig. 1(b) we conclude that we need to choose the smallest possible Bragg angle to enable the largest energy bandwidth as can be estimated also by calculating the differential of the Bragg's law. This conclusion is general and independent of the details of the experiment. In addition, for a fixed lattice spacing, there is a linear dependence between the rocking curve width of Bragg scattering of the beam splitter and the count rate of the heralded photons. For example, for the parameters described above, increasing the rocking curve width by a factor of a hundred leads to an enhancement of the count rate by about 90. We note, however, that the angle of the beam splitter should not be too small to allow the separation between the reflected and transmitted photons.

Of importance, although the mosaic spread deteriorates the reflectivity, it should be sufficiently broad to accommodate the broad angular and spectral distributions of the SPDC process. This tradeoff has to be considered for the design of further x-ray quantum optics experiments with mosaic crystals. Another consideration is the loss in the transmitted beam, which increases when the incident angle of the photons impinging upon the beam splitter decreases. Using a thinner crystal could reduce the absorption but at the expense of the reduction of the reflectivity. For HOPG, the reflectivity decreases significantly below a thickness of 0.2 mm [30].

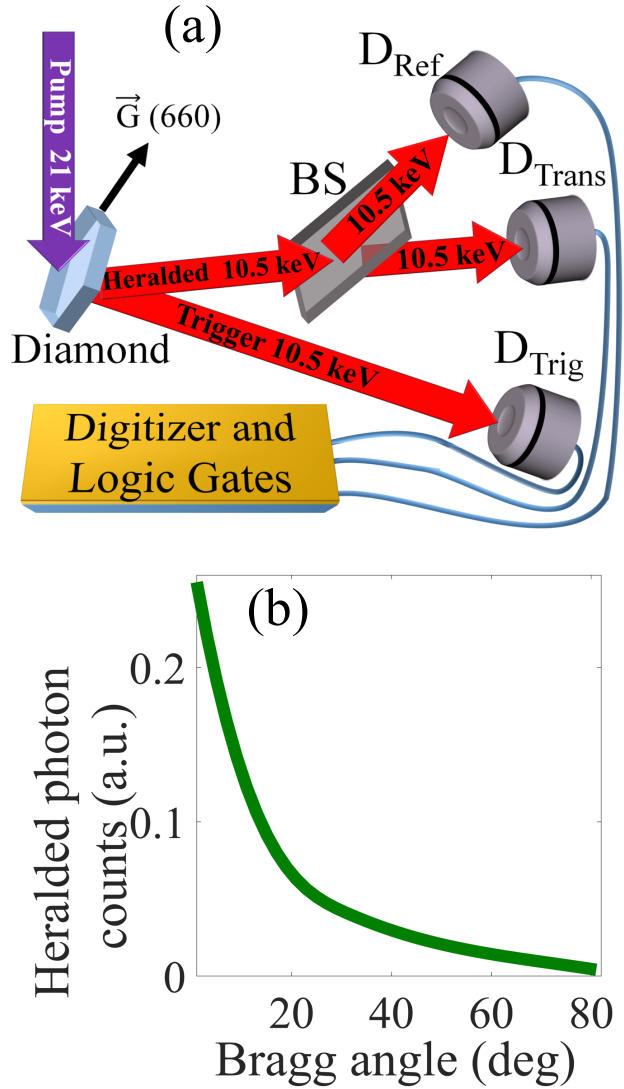

FIG. 1. (a) Experimental setup. The photon pairs are generated in the diamond crystal. The trigger photons are collected by detector $D_{Trig}$ and heralded photons hit the HOPG crystal that is utilized as a Bragg beam splitter (BS). $D_{Trans}$ and $D_{Ref}$ are the detectors for the transmitted and reflected (Bragg scattered) photons, respectively. (b) Simulation results: normalized counts of the heralded photons that are Bragg scattered by the beam splitter as a function of the Bragg angle of the beam splitter. The vertical axis is normalized by the coincidence counts at the output of the SPDC crystal and corrected for absorption in air assuming 10 cm of air path between the SPDC crystal and the detectors.

We performed the experiment at beamline P09 [35] of the PETRA III synchrotron storage ring (DESY, Hamburg). To separate the photon pairs from the background we used logic gates to register only coincidental detection events in which $D_{Trig}$ clicks together with either $D_{Trans}$ or $D_{Ref}$. The time window of the coincidence recording was about 800 ns (expect for the results in Fig. 3 – see details below). To distinguish the down-converted pairs from accidental coincidence counts we post-selected photons according to their energies using the photon energy resolving capability of our detectors. We recorded only photons with photon energies in the range from 7 keV to 17 keV and that the sum of their photon energies is within an energy window of 1 keV around the energy of the pump photon in accord with the conservation of energy and the resolution of our system [36].

We first show that the interaction between the heralded photons and the beam splitter is efficient by exploring the count rates of the heralded photons at each of the output ports of the beam splitter. Figures 2(a) and 2(b) show the spectra of the measured heralded photon counts for the reflected and the transmitted photons, respectively. For the comparison we show the measured spectrum of the trigger detector and plot the numerical calculations for the two spectra. The total heralded photon count rates of the reflected and transmitted photons are $n_R$=0.0093±0.0003 photons/s and $n_T$=0.0164±0.0004 photons/s respectively. These rates are only slightly smaller than the heralded photon count rate we measured before we inserted the beam splitter, $n_H$=0.0583±0.0099 photons/s, and are comparable to the measured coincidence rates in the previous experiments with similar input beam parameters where the photon pairs were measured directly after the nonlinear crystal [26,28,34]. **This is a clear indication that the interaction of the heralded photons with the beam splitter is efficient.** For our experimental parameters our model predicts that the ratios between the rates of the reflected and transmitted photons and the rate of the photon pairs in the absence of the beam splitter are $r_{R-Model}$=0.13 and $r_{T-Model}$=0.17, respectively. The ratios we measured $r_R$=0.159±0.027 and $r_T$=0.281±0.048 - are slightly higher, suggesting that the interaction with the beam splitter is more efficient than predicted. However, this discrepancy can be explained by the improvement in the alignment of the detectors between the two measurements and by a nonlinear response of our detectors due to the strong background in the absence of the beam splitter.

Figure 2 also indicates that, as expected, the measured spectrum of the reflected photons is narrower than spectrum of the transmitted photons since they are Bragg reflected and the agreement between the experimental results and the theory is within the experimental uncertainties. The theoretical dip in the curve of the transmitted beam (Fig. 2(b)) is attributed to Bragg scattering at the energy corresponding to the Bragg angle and cannot be seen in the measurements due to the insufficient energy resolution of our setup.

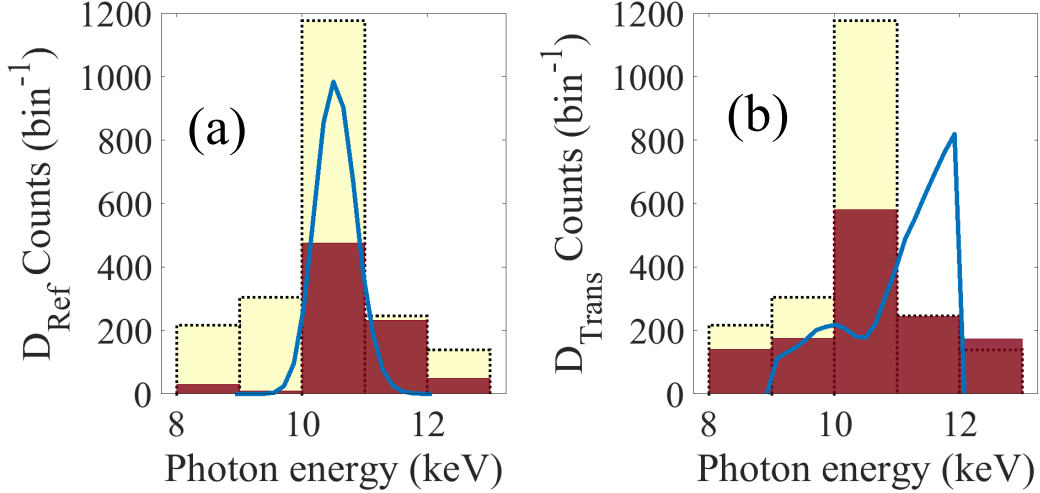

FIG. 2. Photon energy histograms of the counts of the heralded photons at $D_{Ref}$ (a - dark) and $D_{Trans}$ (b - dark) in 88010 seconds and with an energy conservation window of 1 keV. The spectrum of $D_{Trig}$ (light) is shown for the comparison. The blue lines are calculated from theory and scaled vertically to match the total coincidence counts of $D_{Ref}$.

Next, we turn to confirm that the generated radiation is nonclassical. We first show that the correlation between the trigger photons and the photons measured by either $D_{Trans}$ or $D_{Ref}$, within the experiment time window exhibits sub-Poissonian statistics. We calculate the degree of correlation $\sigma \equiv \frac{\langle \delta^2 (N_t - N_h) \rangle}{\langle N_t + N_h \rangle}$, where $\langle \delta^2 x \rangle = \langle x^2 \rangle - \langle x \rangle^2$ is the variance and the average $\langle \ \rangle$ is over the ensemble of detections by $D_{Trig}$ and , $N_t$ and $N_h$, are the number of the trigger photons detected by $D_{Trig}$ and the heralded photons, measured at either $D_{Trans}$ or $D_{Ref}$, respectively. The results plotted in Fig. 3 clearly show that the degree of correlation approaching zero when applying either short time windows or narrow energy windows. This is a conclusive evidence that the generated radiation exhibits sub-Poissonian statistics, hence it is nonclassical. When we open the energy window, the degree of correlation increases with the time window, but it is always smaller than 1. This is because we increase the rate of the accidental coincidences but the probability to measure two photons in the short time window is still low. $\sigma$ decreases also when we narrow the time window but leave the energy conservation window open. When we narrow the energy conservation window, $\sigma$ is nearly zero for any time window we used. See further details on the lowest values of $\sigma$ in the Supplemental Material [33].

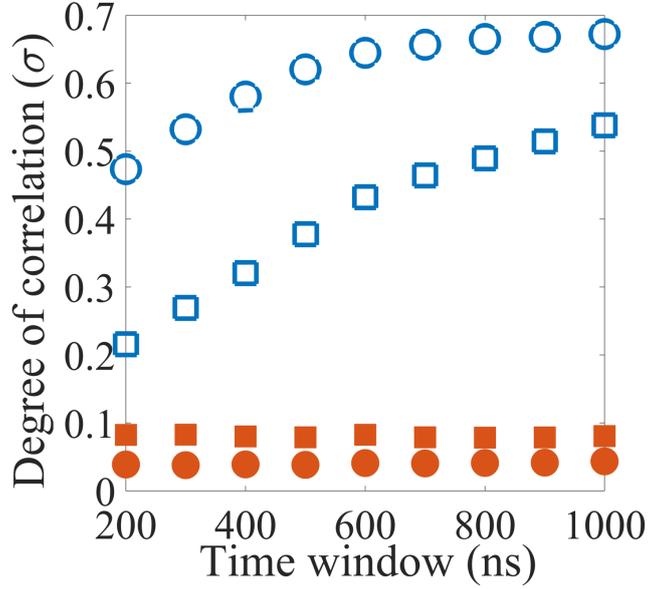

FIG. 3. The degree of correlation versus the coincidence time window for events satisfying the energy conservation within a tolerance of 1 keV (filled circles and rectangles) and for the total events (hollow circles and rectangles). The circles are for $D_{Trans}$ and the squares are for $D_{Ref}$.

Now we turn to show that when the single photons interact with the beam splitter they do not split, as expected. This is one of the most prominent differences between single photons that interact with a beam splitter to classical beams, which are split by the beam splitter. To verify this nonclassical nature of the heralded photons and to ensure that despite the loss in the beam splitter, the quantum nature of the single photons is preserved, we measured the coincidences between the trigger detector and each of the output ports of the beam splitter. **As is clearly seen in Fig. 4 (a), when the energy conservation window is narrow (1 keV), we observe only heralded photons, and we do not measure simultaneous clicks at both outputs of the beam splitter**. **We therefore confirmed that the heralded x-ray photon cannot split.** For the comparison, we show measurements without imposing the photon energy window but for the same number of total counts in Figs. 4(b). Under this condition we measured also accidental coincidences, which are originated from stray radiation. Here we see simultaneous clicks at both outputs, which is an indication that more than one photon interacted with the beam splitter during one detection cycle. To verify that this observation is not fortuitous we show that the number of simultaneous clicks increases with the number of total counts in Fig. 4(c), which represents measurements with the same energy windows as in Fig. 4(b), but the total counts are higher by a factor of 100.

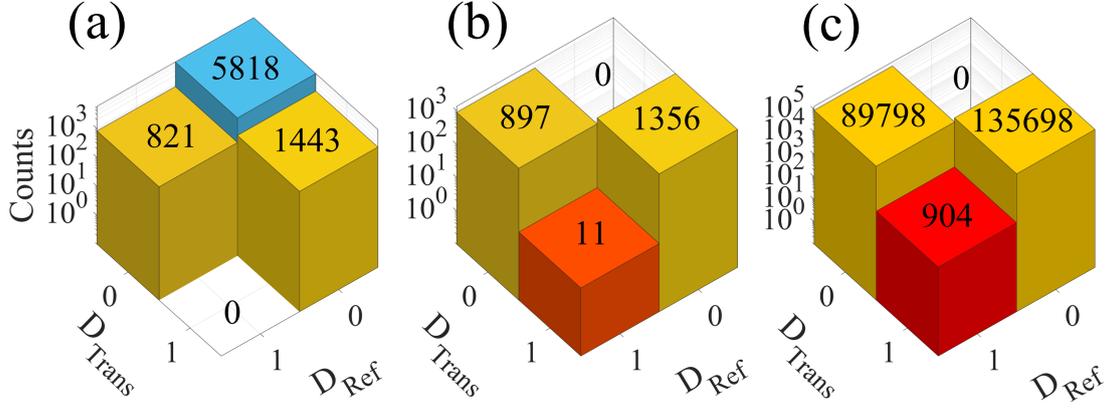

FIG. 4. Count histograms of the photons at the outputs of the beam splitter. In (a) we registered only heralded photons by using photon energy and time filters. In (b) and (c) we registered all the detected photons. In (a) and (b) the total number of events is 2264 and in (c) is 226400. The horizontal axes are the number of counts at each detector in one detection event. The zero-photon column is for events where only the trigger detector detects photons with photon energies in the selected range (since in (b) and (c) the energy window is wide open there are no counts in the zero-photon columns).

To quantify the purity of the quantum state, we use the anticorrelation criterion [8,37,38],

$$\alpha = \frac{N_{Trig} N_{Trig-T-R}}{N_{Trig-T} N_{Trig-R}} \quad (1)$$

Here $N_{Trig}$ is the total number of trigger events, $N_{Trig-T}$ and $N_{Trig-R}$ are the number of coincidences between $D_{Trans}$ and $D_{Ref}$, respectively with $D_{Trig}$. $N_{Trig-T-R}$ is the number of triple coincidences between $D_{Trans}$ and $D_{Ref}$ and $D_{Trig}$. $N_{Trig}$ as the number of events in which $D_{Trig}$ and at least one of the detectors $D_{Trans}$ or $D_{Ref}$ measure photons within a predefined energy window for each detector. According to this criterion, for single photons, $\alpha$ is smaller than 1 while for classical beams is larger than 1.

For the heralded photons (Fig. 4(a)) we found that $\alpha$ is nominally zero, which is the indication of background-free quantum behavior. This is in contrast to most analog quantum optics experiments in the visible range in which $\alpha$ is smaller than 1 but finite [39,40]. This is a clear demonstration of the ability to perform background-free quantum optics experiments with x-rays.

Interestingly, $\alpha$ is smaller than 1 even when most of the detected photons are originated from stray radiation. The reason is that even with this radiation during a single measurement interval, only one photon interacts with the beam splitter on average and the probability that two photons interact with the beam splitter is much lower. This is because we use short coincidence time windows to reduce the background in our experiments. Consequently, since a single photon is a single photon that cannot split regardless its origin, at most events there will be no simultaneous clicks at both output ports of the beam splitter leading to $\alpha < 1$. However, there is always a small probability that two simultaneous photons arrive, hence for the stray light $\alpha$ is not zero (for example $\alpha$ is $0.02\pm0.006$ and $0.0168\pm0.0006$ for Figs. 4(b) and 4(c), respectively. The data we used for the calculation are given in the Supplemental Material [33]). Our results highlight that the anticorrelation criterion does not imply that every photon we measured was a single photon but only that on average we measured single photons.

In summary, this work reports the direct evidence that x-ray photons are undividable quanta and the proof of principle experiment demonstrating efficient interaction of x-ray single photons with a beam splitter. The heralded photon rates we have observed at the output ports of the beam splitter are comparable to previously reported count rates in experiments when the photons have been measured directly after the generating crystal [26,34]. Further improvements of the efficiency can be obtained by improving the match between the angular dispersion of the Bragg scattering of the beam splitter and the angular dispersion of the SPDC. This can be done by tuning the phase matching angles of the SPDC and by choosing a small Bragg angle and broad angular acceptance for the beam splitter. The single photon statistics we have observed exhibit high fidelity despite the existence of loss and background noise in the setup. Our work opens new possibilities for x-ray quantum optics by enabling experiments, which rely on beam splitters and single photon interactions. Further generalization of our work can lead to the development of novel sensitive and precise measurement techniques based upon x-ray single photon interferometry or NOON x-ray states.

This research was carried out at beamline P09 at PETRA III at DESY, a member of the Helmholtz Association (HGF). We would like to thank David Reuther for assistance during the experiment. The research leading to this result has been supported by the project CALIPSOplus under the Grant Agreement 730872 from the EU Framework Programme for Research and Innovation HORIZON 2020. This work was supported by the Israel Science Foundation (ISF), Grant No. 201/17. The authors thank Eliahu Cohen for helpful discussions.

E-mail me at: Sharon.shwartz@biu.ac.il

**References**

[1] M. O. Scully and M. S. Zubairy, *Quantum Optics* (Cambridge University Press, 1997).

[2] Y. Yamamoto and A. Imamoglu, *Mesoscopic Quantum Optics* (John Wiley & Sons, New York, 1999).

[3] D. F. Walls and G. J. Milburn, *Quantum Optics* (Springer-Verlag Berlin Heidelberg, Berlin, 1995).

[4] D. Bouwmeester, A. K. Ekert, and A. Zeilinger, *The Physics of Quantum Information: Quantum Cryptography, Quantum Teleportation, Quantum Computation* (Springer, New York, 2000).

[5] C. K. Hong, Z. Y. Ou, and L. Mandel, Phys. Rev. Lett. **59**, 2044 (1987).

[6] A. C. Elitzur and L. Vaidman, Found. Phys. **23**, 987 (1993).

[7] A. G. White, J. R. Mitchell, O. Nairz, and P. G. Kwiat, Phys. Rev. A **58**, 605 (1998).

[8] P. Grangier, G. Roger, and A. Aspect, Europhys. Lett. **1**, 173 (1986).

[9] T. Jennewein, C. Simon, G. Weihs, H. Weinfurter, and A. Zeilinger, Phys. Rev. Lett. **84**, 4729 (2000).

[10] P. Kok, H. Lee, and J. P. Dowling, Phys. Rev. A **65**, 52104 (2002).

[11] R. Röhlsberger, J. Evers, and S. Shwartz, in *Synchrotron Light Sources Free. Lasers Accel. Physics, Instrum. Sci. Appl.*, edited by E. J. Jaeschke, S. Khan, J. R. Schneider, and J. B. Hastings (Springer International Publishing, Cham,

2020), pp. 1399–1431.

[12] M. I. Kolobov, *Quantum Imaging* (Springer, New York, 2007).

[13] G. Brida, M. Genovese, and I. Ruo Berchera, Nat. Photonics **4**, 227 (2010).

[14] P. A. Morris, R. S. Aspden, J. E. C. Bell, R. W. Boyd, and M. J. Padgett, Nat. Commun. **6**, 5913 (2015).

[15] S. Lloyd, Science **321**, 1463 (2008).

[16] M. J. Holland and K. Burnett, Phys. Rev. Lett. **71**, 1355 (1993).

[17] V. Giovannetti, S. Lloyd, and L. Maccone, Science **306**, 1330 (2004).

[18] M. W. Mitchell, J. S. Lundeen, and A. M. Steinberg, Nature **429**, 161 (2004).

[19] L. Pezzé, A. Smerzi, G. Khoury, J. F. Hodelin, and D. Bouwmeester, Phys. Rev. Lett. **99**, 223602 (2007).

[20] G. Y. Xiang, B. L. Higgins, D. W. Berry, H. M. Wiseman, and G. J. Pryde, Nat. Photonics **5**, 43 (2011).

[21] V. Giovannetti, S. Lloyd, and L. Maccone, Nat. Photonics **5**, 222 (2011).

[22] B. W. Adams, *Nonlinear Optics, Quantum Optics, and Ultrafast Phenomena with X-Rays: Physics with X-Ray Free-Electron Lasers* (Kluwer Academic Publisher, Norwell, MA, 2008).

[23] S. Shwartz and S. E. Harris, Phys. Rev. Lett. **106**, 80501 (2011).

[24] A. Pálffy, C. H. Keitel, and J. Evers, Phys. Rev. Lett. **103**, 17401 (2009).

[25] F. Vagizov, V. Antonov, Y. V Radeonychev, R. N. Shakhmuratov, and O. Kocharovskaya, Nature **508**, 80 (2014).

[26] S. Shwartz, R. N. Coffee, J. M. Feldkamp, Y. Feng, J. B. Hastings, G. Y. Yin, and S. E. Harris, Phys. Rev. Lett. **109**, 13602 (2012).

[27] A. Schori, D. Borodin, K. Tamasaku, and S. Shwartz, Phys. Rev. A **97**, 63804 (2018).

[28] S. Sofer, E. Strizhevsky, A. Schori, K. Tamasaku, and S. Shwartz, Phys. Rev. X **9**, 31033 (2019).

[29] A. Authier, *Dynamical Theory of X-Ray Diffraction* (Oxford University Press, New York, 2001).

[30] A. K. Freund, A. Munkholm, and S. Brennan, in *Opt. High-Brightness Synchrotron Radiat. Beamlines II*, edited by L. E. Berman and J. Arthur (SPIE, 1996), pp. 68–79.

[31] D. Attwood, *Soft X-Rays and Extreme Ultraviolet Radiation: Principles and Applications* (Cambridge University Press, Cambridge, 1999).

[32] S. M. Barnett, J. Jeffers, A. Gatti, and R. Loudon, Phys. Rev. A **57**, 2134 (1998).

[33] See Supplemental Material at URL for further details on the theoretical model,

the electronics used for the experiment, and background levels for the experimental measurements.

[34] D. Borodin, A. Schori, F. Zontone, and S. Shwartz, Phys. Rev. A **94**, 013843 (2016).

[35] J. Strempfer, S. Francoual, D. Reuther, D. K. Shukla, A. Skaugen, H. Schulte-Schrepping, T. Kracht, and H. Franz, J. Synchrotron Radiat. **20**, 541 (2013).

[36] We collected also data with wider energy ranges, which is presented in Fig. 3 and Fig. 4. We used various energy ranges in order to compare the heralded photons with the classical beam.

[37] A. B. U'Ren, C. Silberhorn, J. L. Ball, K. Banaszek, and I. A. Walmsley, Phys. Rev. A **72**, 21802 (2005).

[38] M. Beck, J. Opt. Soc. Am. B **24**, 2972 (2007).

[39] J. Zhao, C. Ma, M. Rüsing, and S. Mookherjea, Phys. Rev. Lett. **124**, 163603 (2020).

[40] M. D. Eisaman, J. Fan, A. Migdall, and S. V Polyakov, Rev. Sci. Instrum. **82**, 71101 (2011).

[41] R. Loudon, *The Quantum Theory of Light* (Oxford University Press, New York, 2000).

# Supplemental material for: "Efficient Interaction of Heralded X-ray Photons with a Beam Splitter"

E. Strizhevsky[1], D. Borodin[1], A. Schori[1,2], S. Francoual[3], R. Röhlsberger[3], and S. Shwartz[1*]

[1]*Physics Department and Institute of Nanotechnology, Bar-Ilan University, Ramat Gan, 5290002 Israel*

[2]*PULSE Institute, SLAC National Accelerator Laboratory, Menlo Park, California 94025, USA*

[3]*Deutsches Elektronen-Synchrotron DESY, Notkestrasse 85, D-22607 Hamburg, Germany*

## Positions of the diamond crystal and detectors:

Here we provide further details on our experimental setup. The theoretical value of the Bragg angle of the C(660) atomic planes is 44.61°. For the generation of the photon pairs, we rotate the angle of the diamond crystal by 0.008° from the Bragg angle and set the angles of the trigger detector and the beam splitter with respect to the diamond atomic planes to 43.63° and 45.59°, respectively, according to the phase matching condition. The distance between the detectors and the diamond crystal is $1000 \pm 10$ mm where we use a helium tube of $900 \pm 10$ mm length and $200 \pm 5$ mm diameter to reduce air absorption and scattering. The synchrotron beam dimensions are about 2 mm and 0.2 mm in the vertical and horizontal directions, respectively.

## Calculation of interaction between the heralded photons and the beam splitter:

In this section we provide further details on our procedure for the calculation of heralded photon rate at the output ports of the beam splitter. Since the count rates of heralded photons are actually the coincidence count rates between the heralded photons and the trigger photons, we need to calculate the coincidence count rate between each of the output ports of the beam splitter and the trigger detector. We use the second order Glauber correlation function, which is given by $R_c = S \iint \langle a_{Trig}^\dagger(\mathbf{r}_2, t_2) a_{Heral}^\dagger(\mathbf{r}_1, t_1) a_{Heral}(\mathbf{r}_1, t_1) a_{Trig}(\mathbf{r}_2, t_2) \rangle d\mathbf{u} d\tau$ where $S$ is the area of the pump at the input of the nonlinear crystal, $\mathbf{u} = \mathbf{r}_2 - \mathbf{r}_1$, and $\tau = t_2 - t_1$ [26]. Since it is more convenient to calculate the frequency domain operators than the time-space operators we use the relation $a_j(z,\mathbf{r},t) = \int_0^\infty \int_{-\infty}^\infty a_j(z,\mathbf{q},\omega) \exp[-i(\mathbf{q}\cdot\mathbf{r} - \omega t)] d\mathbf{q} d\omega$, where $\mathbf{r} = (x,y)$ and $\mathbf{q} = (k_x, k_y)$ to transfrom the operator to the frequncy domain (in time and in space). The relation between $\omega$, the photon angular frequency, and the magnitude of the wave-vector is $\omega_j = k_j c / n(\omega_j)$ and the wave-vector components $k_x$ and $k_y$ are parallel to the surfaces of the nonlinear crystal. The operators satisfy the commutation relations -

$$[a_j(z_1,\mathbf{q}_1,\omega_1), a^\dagger_k(z_2,\mathbf{q}_2,\omega_2)] = \frac{1}{(2\pi)^3} \delta_{j,k} \delta(z_1 - z_2) \delta(\mathbf{q}_1 - \mathbf{q}_2) \delta(\omega_1 - \omega_2).$$

To calculate the frequency domain heralded and trigger operators we first calculate their values at the output of the nonlinear crystal by solving the lossless coupled equations assuming the undepleted pump approximation in the Heisenberg picture

$$\frac{\partial a_{Heral}}{\partial z} = \kappa a^\dagger_{Trig} \exp(i\Delta k_z z)$$
$$\frac{\partial a^\dagger_{Trig}}{\partial z} = \kappa^* a_{Heral} \exp(-i\Delta k_z z)$$
(1)

Here $\kappa$ is the coupling constant that includes the nonlinear coefficient and the pump intensity and $\Delta k_z = k_p \cos(\theta_p) - k_{Heral} \cos(\theta_{Heral}) - k_{Trig} \cos(\theta_{Trig})$ is the phase mismatch along the z direction. $\theta_p, \theta_{Heral}$, and $\theta_{Trig}$ are the angles between the atomic planes and the wave vectors of the pump, heralded, and trigger photons, respectively.

Next, we need to incorporate the expression that presents the beam splitter. The aim of this numerical simulation is to examine how the parameters of the mosaic crystal impact its efficiency as a beam splitter for the down converted photons, which possess broad energy and angular distributions. However, there is no simple analytical expression for the reflection coefficient for mosaic crystals [29], and we want the model to be as clear as possible, so that the important parameters could be easily identified and characterized. Therefore, we use a Gaussian model for the reflection coefficient and neglect the additive quantum noise from the open port of the beam splitter. The latter is justified since we are interested in the mean output photon numbers of the heralded photons (the correlation between the trigger and one of the output ports of the beam splitter) and not trying to calculate the correlation between the output ports of the beam splitter [41]. The Gaussian function that we choose for our model is -

$$R(\omega_{Heral}, \Delta\theta) = \sqrt{A} \exp\left\{-\frac{1}{2}\left[\frac{\Delta\theta + \theta_B\left(\frac{1}{2}\omega_p\right) - \theta_B(\omega_{Heral})}{b}\right]^2\right\} \qquad (2)$$

This model incorporates the important parameters and describes reasonably the dependence of the reflectance of the mosaic crystal on the deviation from the Bragg angle and its spectral dependence. Here the frequency dependence is originated from the Bragg's law for a given incident beam frequency $\theta_B(\omega_{Heral}) = \sin^{-1}\left(\frac{\pi c}{d\omega_{Heral}}\right)$, where c is the speed of light and d is the lattice interplanar spacing. The angular deviation $\Delta\theta$ is defined relative to the Bragg angle at the heralded photon wavelength as described in Fig. 1. The peak reflectivity of the HOPG is denoted by A and its value is taken to be 0.5 [30]. The width parameter b is 0.48° deduced form the measured Full Width at Half Maximum (FWHM) of the rocking curve of the HOPG, which is 0.8°.

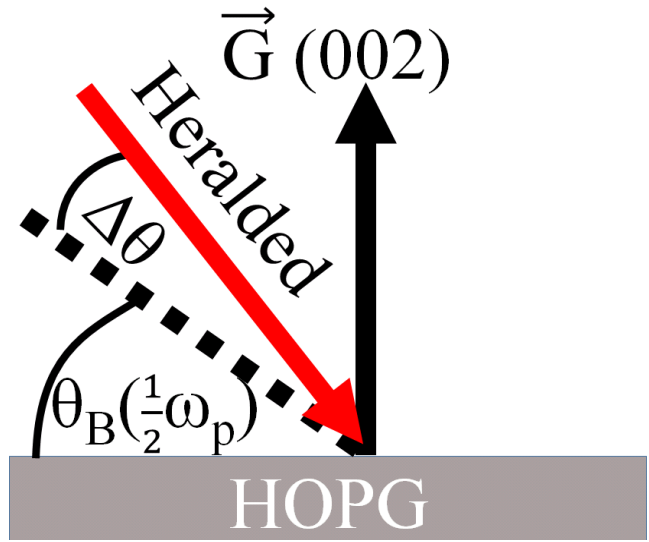

FIG. 1. Description of the angles of the beam that hits the beam splitter.

Finally, we multiply the heralded operator we found by solving Eq. 1 by the expression in Eq. 2 and calculate the coincidence count rate by numerically integrating over photon energies in the range of 9.5 keV to 11.5 keV and an angular range of 5 mrad centered at the phase matching angle, which covers an area of about 20 mm$^2$ on the detector.

### Coincidence electronics and data acquisition

Here we provide further details on the coincidence electronics, which we use to verify that the photon pairs arrive simultaneously and to reduce background radiation. Our detectors generate two types of signals for each detected photon: analog voltage pulses with height that is proportional to the photon energy of the detected photons and logic pulses with a fixed height of 1.4 V. The pulse duration of the analog signal is 200 ns and the pulse duration of the logic signal is 1000 ns. The logic pulse is generated only when a photon within a predefined energy range is detected (functions as an output of a single channel analyzer).

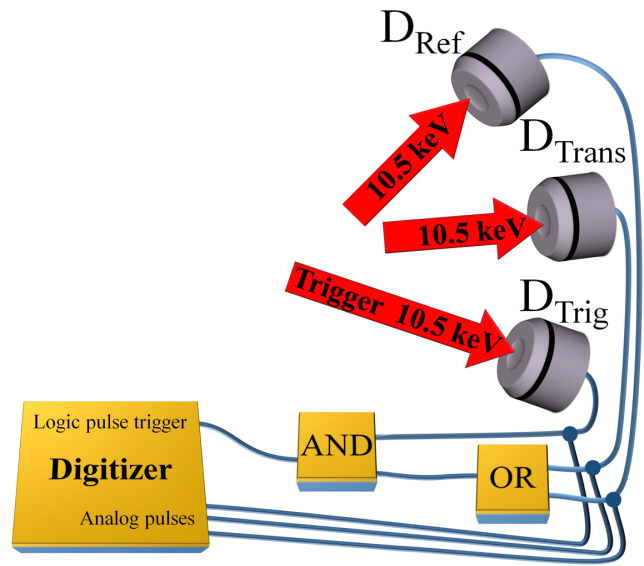

FIG. 2. Schematic of the coincidence electronics. See details in the text.

As can be seen in Fig. 2 we use logic gates to trigger a digitizer when the logic pulses from detector $D_{Trig}$ and at least one of the detectors $D_{Trans}$ or $D_{Ref}$ overlap. The overlap point is determined by the beginning of overlap between two logic pulses. We used logic gates triggering since otherwise the raw count rate of the detector would lead to overflow of the buffer of our digitizer. These logic gates reduce the number of the event rates that are registered by our digitizer to less than 200 events per second. After the measurement we scan the data and use a software filter to register only events that their analog pulses are within a time window of ±800 ns around the overlap point. This procedure improves the signal-to-noise ratio as can be seen from our background free results presented in the main text. However, if two photons are separated by 800-1000 ns they still can trigger the digitizer (due to the length of the logic pulses), but one of them is partially outside the time window of our software filter, which restricts the detection window to 1600 ns. In this case only the photon that is in the window allowed by the software is registered, thus the system registers an event with only one photon and not a pair of photons. We can of course use the software to filter out events that are not the detection of pairs of photons, but we used it to demonstrate the reduction in $\sigma$ as it is shown in Fig. 3 of the main text.

The functionally of our electronics explain also why the degree of correlation, $\sigma$ presented in Fig. 3 of the main text is not identical to zero even when we use the stringent conditions for time windows and energy conservation. In order to show the gradual decrease of $\sigma$ with the time window, we included detection events where the peak of one of the analog pulses was outside the maximal time window we set for the software filter, as shown in Fig. 3, which is concidered as a detection of only one photon (detector D$_{Trig}$ in this example). These single detections can be within the energy conservation window if the energy of these photons is close to the energy of the pump photon. Such events contribute a non zero values to the average calculation at the numerator of $\sigma$.

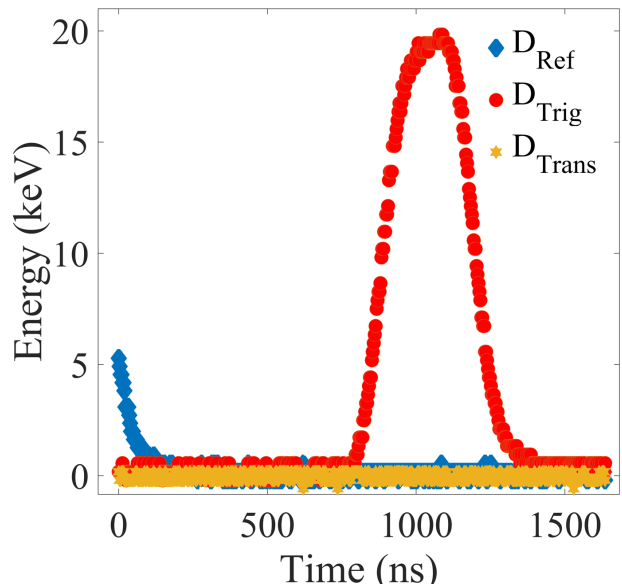

FIG. 3. An example for measured digitizer traces when only detector $D_{Trig}$ was recorded. In this example detector $D_{Trig}$ detected a photon at about 20 keV. Detector $D_{Ref}$ also detected a photon in the photon energy window allowed by our system and the digitizer was triggered to record the analog signal. However, the temporal separation between the two analog signals was larger than 800 ns, and the analog signal of detector $D_{Ref}$ was not registered.

**Calculating $\alpha$ from the experimental results in Fig. 4(b) and Fig. 4(c) of the main text:**

Finally, we provide the measured counts that we used for the calculation of $\alpha$ :

| Figure | $N_{Trig}$ | $N_{Trig-T}$ | $N_{Trig-R}$ | $N_{Trig-T-R}$ | $\alpha$ |
|---|---|---|---|---|---|
| Fig. 4(b) | 2264 | 897 | 1356 | 11 | $0.02 \pm 0.006$ |
| Fig. 4(c) | 226400 | 89798 | 135698 | 904 | $0.0168 \pm 0.0006$ |